\documentclass[12pt]{article}
\usepackage{graphics}
\usepackage[dvips]{graphicx}
\usepackage{amssymb}
\usepackage{amsmath}
\newcommand{\beq}{\begin{equation}}
\newcommand{\eeq}{\end{equation}}
\newcommand{\bea}{\begin{eqnarray}}
\newcommand{\eea}{\end{eqnarray}}

\def\e2sig{e^{-2r\sigma}}

\makeatletter
\@addtoreset{equation}{section}

\makeatother

\begin{document}
\setlength{\baselineskip}{18pt}

\begin{titlepage}

\vspace*{2cm}

\begin{center}
{\Large\bf Direct Gauge Mediation of \\
\vspace*{3mm}
Uplifted Metastable Supersymmetry Breaking \\
\vspace*{3mm}
in Supergravity
} 
\end{center}
\vspace{20mm}

\begin{center}
{\large
Nobuhito Maru
}
\end{center}
%
%
%
%
\begin{center}
{\it Department of Physics, Chuo University,
Tokyo 112-8551, Japan}
\end{center}
%
%
\vspace*{3cm}

\centerline{\large\bf Abstract}
\vspace*{1cm}
We propose a direct gauge mediation model based on 
an uplifted metastable SUSY breaking coupled to supergravity. 
A constant superpotential plays an essential role to fix the moduli 
as well as breaking SUSY and R-symmetry and the cancellation of the cosmological constant. 
Gaugino masses are generated at leading order of SUSY breaking scale, 
and comparable to the sfermion masses as in the ordinary gauge mediation. 
Landau pole problem for QCD coupling can be easily solved 
since more than half of messengers become superheavy, which are heavier than the GUT scale. 
\end{titlepage}

\section{Introduction}

Supersymmetry (SUSY) is one of the fascinating scenarios solving the hierarchy problem. 
However, it has to be broken at low energy to be relevant to nature. 
According to Witten index argument \cite{Witten}, 
SUSY can be broken nonperturbatively in chiral gauge theories 
except for the vector-like model in the special case 
with the color and flavor numbers \cite{IYIT}. 
Model building by use of chiral gauge theories is not so easy 
although the nonperturbative dynamics of SUSY theories has been clarified by Seiberg \cite{Seiberg}. 
Several years ago, Intriligator, Seiberg and Shih (ISS) have discovered 
a metastable SUSY breaking vacuum in (light) massive SUSY QCD 
in a free magnetic phase \cite{ISS}. 
A remarkable fact is that the models they proposed are vector-like models 
in a wide range of the number of flavors larger than the number of the color. 
For SUSY breaking model builders, this fact immediately leads to the idea 
that the Landau pole problem of QCD coupling, 
which was very hard to solve for a long time, can be easily solved. 
However, the ISS model has a basic problem that an R-symmetry is not broken 
in their vacuum, so the (Majorana) gaugino masses cannot be generated 
after the mediation of SUSY breaking. 
In order to overcome this problem, 
there have been many works \cite{many} from various viewpoints  
by the way to break an R-symmetry explicitly or spontaneously. 

Once SUSY and R-symmetry breaking are realized, 
the next task is to transmit its breaking to our world. 
Of various mediation mechanisms, 
the direct gauge mediation \cite{DGM} is phenomenologically attractive 
since there is no FCNC problem and it is economical from the model building viewpoint. 
However, only breaking R-symmetry is not enough and 
the anomalously small gaugino mass problem has been arisen for 
the proposed direct gauge mediation models 
\cite{many}.\footnote{This problem has already been recognized before in \cite{INTY}.} 
Gaugino masses are not generated at linear order of SUSY breaking scale, 
but at least generated at the third order. 
This immediately leads to the fine-tuning in the Higgs mass 
since the sfermion masses become heavy comparing to the gaugino masses 
as in the split SUSY scenario \cite{AD}. 
This problem has been recently studied in \cite{KS} 
and the authors have shown that the smallness of gaugino masses is related to 
the global structure of vacua in renormalizable theories. 
In particular, it has been shown that models which breaks SUSY in the lowest energy state 
necessarily lead to anomalously small gaugino masses. 
To avoid this situation, namely to obtain the gaugino masses at leading order of SUSY breaking, 
SUSY must be broken in uplifted vacua. 
Along this direction, several interesting models has been known 
\cite{KOO, GKK, AJK, Barnard, AEG, HO}. 

In this paper, we propose a direct gauge mediation model 
based on an uplifted SUSY breaking \cite{GKK} coupled to supergravity (SUGRA). 
The model of \cite{GKK} deforms not only a moduli space 
such that the rank of the dual quark bilinears is not maximal, 
but also the mass terms of quarks in an ISS model. 
Then, SUSY breaking vacuum is uplifted by the deformation and 
the gaugino masses comparable to the sfermion masses are generated 
as in the ordinary gauge mediation \cite{GM}. 
The model is quite interesting but has some unappealing points as discussed in \cite{GKK}. 
The first point is that an explicit R-symmetry breaking term of the moduli has to be introduced 
since the moduli is undetermined as it stands, 
and what is worse the overall coefficient is required to be tiny 
so that the potential have the minimum.  
The second point is that the perturbative coupling unification is difficult to achieve 
since the number of messengers is large. 
This is a similar situation as the direct gauge mediation using the chiral gauge theories
These drawbacks are solved in our model by only coupling an uplifted SUSY breaking model \cite{GKK} 
to supergravity as follows. 
R-symmetry breaking is realized by introducing a nonzero constant superpotential with R-charge 2. 
As a result, the moduli is lifted by the constant superpotential. 
The magnitude of the constant superpotential is determined 
by canceling the cosmological constant. 
Moreover, the Landau pole problem is easily avoided 
because many of the messengers becomes superheavy, 
which are larger than the GUT scale. 

This paper is organized as follows. 
In the next section, we discuss our model.
We will see that a constant superpotential plays an essential role 
to lift the moduli as well as canceling the cosmological constant 
and the breaking of SUSY and R-symmetry. 
The messenger mass spectrum is calculated in section 3, 
which is needed in the Landau pole analysis. 
In section 4, the direct gauge mediation is discussed. 
We will show that Landau pole problem can be avoided 
in a wide range of parameters in section 5. 
The last section 6 summarizes the results of our paper. 

\section{Model}

In this paper, we consider an uplifted metastable SUSY breaking model \cite{GKK} 
of Intriligator-Seiberg-Shih (ISS) \cite{ISS} coupled to supergravity (SUGRA). 
The uplifted ISS model is a deformed model of an 
${\cal N}=1$ SUSY $SU(N)(N \equiv N_f-N_c)$ gauge theory 
with $N_c+1 < N_f < \frac{3}{2}N_c$ flavor quarks and the gauge singlet couples to the quarks 
in the superpotential. 
The matter content and symmetries are summarized as follows. 
\bea
\begin{tabular}{ccccccc}
\hline
 & $SU(N)$ & $SU(N_f)$ & $SU(N_f)$ & $U(1)_B$ & $U(1)$ & $U(1)_R$ \\
\hline
$q$ & ${\bf N}$ & ${\bf {\bar{N}_f}}$ & 1 & 1 & 1 & 0 \\
$\tilde{q}$ & ${\bf \bar{N}}$ & ${\bf 1}$ & ${\bf N_f}$ & $-$1 & 1 & 0 \\
$\Phi$ & ${\bf 1}$ & ${\bf N_f}$ & ${\bf \bar{N}_f}$ & 0 & $-$2 & 2 \\
\hline
\end{tabular}
\eea
where $SU(N)$ is a gauge group and others are global symmetries. 
The superpotential and K\"ahler potential are given by
\bea
W &=& h q \Phi \tilde{q} - h \mu_1^2 \sum_{i=1}^k \Phi^i_i 
- h \mu_2^2 \sum_{i=k+1}^{N_f} \Phi^i_i + c, 
\label{spot} \\
K &=& {\rm Tr}[\Phi^\dag \Phi + q^\dag q + \tilde{q}^\dag \tilde{q} + \cdots].  
\label{kahler}
\eea
where the trace is taken in flavor space. 
$c$ is a constant superpotential necessary 
for the cancellation of the cosmological constant. 
The ellipsis in the K\"ahler potential means higher dimensional terms. 
Note that this model is dual to ${\cal N}=1$ $SU(N_c)$ SUSY gauge theory 
with $N_f$ flavors of quarks $Q$ in the range $N_c+1 < N_f <\frac{3}{2}N_c$ 
and the superpotential
\bea
W = m_1 \sum_{i=1}^{k} Q^i \bar{Q}_i 
+ m_2 \sum_{i=k+1}^{N_f} Q^i \bar{Q}_i. 
\eea
The mass scales are related as $\mu_{1,2}^2 = -m_{1,2}\Lambda$ 
where $\Lambda$ is a dynamical scale of the $SU(N_c)$ gauge theory.

Let us first study the vacuum structure of the theory. 
The classical flat directions are parameterized as follows. 
\bea
\Phi = 
\left(
\begin{array}{cc}
V_{k \times k} & Y_{k \times (N_f-k)} \\
\tilde{Y}_{(N_f-k) \times k} & Z_{(N_f-k) \times (N_f-k)} \\
\end{array}
\right) 
&\to& 
\left(
\begin{array}{cc}
{\bf 0}_{k \times k} & {\bf 0}_{k \times (N_f-k)} \\
{\bf 0}_{(N_f-k) \times k} & Z_{(N_f-k) \times (N_f-k)} \\
\end{array}
\right), \\
q = 
\left(
\begin{array}{cc}
(\chi_1)_{k \times k} & (\rho_1)_{k \times (N_f-k)} \\
(\chi_2)_{(N-k) \times k} & (\rho_2)_{(N-k) \times (N_f-k)} \\
\end{array}
\right)
&\to& 
\left(
\begin{array}{cc}
\mu_1 {\bf 1}_{k \times k} & {\bf 0}_{k \times (N_f-k)} \\
{\bf 0}_{(N-k) \times k} & {\bf 0}_{(N-k) \times (N_f-k)} \\
\end{array}
\right), \\
\tilde{q} = 
\left(
\begin{array}{cc}
(\tilde{\chi}_1)_{k \times k} & (\tilde{\chi}_2)_{k \times (N-k)} \\
(\tilde{\rho}_2)_{(N_f-k) \times k} & (\tilde{\rho}_2)_{(N_f-k) \times (N-k)} \\
\end{array}
\right)
&\to& 
\left(
\begin{array}{cc}
\mu_1 {\bf 1}_{k \times k} & {\bf 0}_{k \times (N-k)} \\
{\bf 0}_{(N_f-k) \times k} & {\bf 0}_{(N_f-k) \times (N-k)} \\
\end{array}
\right)
\eea
where $k \le N$ \cite{GKK}. 
For $k=N$, the ISS model is recovered. 
Using the degrees of freedom of the global symmetries $SU(k) \times SU(N_f-k)$, 
we find the classical solution as in the right-hand side. 
The vacuum expectation values (VEVs) of $\chi_1, \tilde{\chi}_1$ will be determined later 
by the F-flatness condition.  

The VEV of the above fields are determined by the minimization of the scalar potential. 
In the classical flat directions, the superpotential and the K\"ahler potential are decomposed as
\bea
K &=& {\rm Tr}(|V|^2 + |\tilde{Y}|^2 + |Z|^2 + |Y|^2 
+ |{\chi_1}|^2 + |{\chi_2}|^2 + |\tilde{\chi}_1|^2 + |{\tilde{\chi}_2}|^2) 
\nonumber \\ 
&&+ {\rm Tr}(|{\rho_1}|^2 + |{\rho_2}|^2 + |{\tilde{\rho}_1}|^2  
+ |{\tilde{\rho}_2}|^2 + \cdots), \\
W &=& -h \mu_2^2 {\rm Tr}Z +h \sum_{i}^{N-k} Z \rho_2^i (\tilde{\rho}_2)_i 
+ h \sum_i^k \left[ Z \rho_1^i (\tilde{\rho}_1)_i + \mu_1 \rho_1^i \tilde{Y}_i 
+ \mu_1 Y^i (\tilde{\rho}_1)_i \right] + c.  
\eea
It is useful to combine the K\"ahler potential and the superpotential 
into a single form
\bea
G = K + \ln|W|^2 
\eea
where the Planck scale is set to be the unity. 
Throughout this paper, this convention is understood.

Using this function $G$, the scalar potential in supergravity is given by
\bea
V &=& e^G 
{\rm Tr}
\left[
G^{VV^\dag} |G_V|^2 + G^{ZZ^\dag} |G_Z|^2 + G^{YY^\dag} |G_Y|^2 
+ G^{\tilde{Y}\tilde{Y}^\dag} |G_{\tilde{Y}}|^2 
\right. \nonumber \\
&& \left. 
+ G^{\chi_1 \chi_1^\dag} |G_{\chi_1}|^2 + G^{\chi_2 \chi_2^\dag} |G_{\chi_2}|^2  
+ G^{\tilde{\chi}_1 \tilde{\chi}_1^\dag} |G_{\tilde{\chi}_1}|^2 
+ G^{\tilde{\chi}_2 \tilde{\chi}_2^\dag} |G_{\tilde{\chi}_2}|^2
\right. \nonumber \\
&& \left. 
+ G^{\rho_1 \rho_1^\dag} |G_{\rho_1}|^2 + G^{\rho_2 \rho_2^\dag} |G_{\rho_2}|^2
+ G^{\tilde{\rho}_1 \tilde{\rho}_1^\dag} |G_{\tilde{\rho}_1}|^2 
+ G^{\tilde{\rho}_2 \tilde{\rho}_2^\dag} |G_{\tilde{\rho}_2}|^2 - 3
\right] \nonumber \\
&=& e^K 
{\rm Tr}
\left[
|h (\chi_1 \tilde{\chi}_1 - \mu_1^2)|^2 + |Z^\dag W - h \mu_2^2|^2 
+ |\chi_1^\dag W|^2 + |\tilde{\chi}_1^\dag W|^2 - 3|W|^2
\right] 
\label{pot}
\eea
where $G_{\phi_i\phi^\dag_j} \equiv \frac{\partial^2 G}{\partial \phi_i \partial \phi^*_j}$.

SUSY vacuum conditions are given by
\bea
0 &=& \chi_1 \tilde{\chi}_1 -\mu_1^2, \\
0 &=& Z^\dag W - h \mu_2^2 = Z^\dag (-h\mu_2^2 {\rm Tr}Z +c) - h \mu_2^2, \\
0 &=& W = -h\mu_2^2 {\rm Tr}Z + c,
\eea
which are obviously incompatible, namely SUSY is broken.
The vacuum energy is 
\bea
V_0 = {\rm Tr}|F_Z|^2
= (N_f-k)|h \mu_2^2|^2
\eea
where $V_0$ means the leading terms of the potential independent of 
the Planck scale.\footnote{Throughout this paper 
$e^K \simeq e^{|\langle Z \rangle|^2} \simeq 1$ is taken by (\ref{trZ}).} 
At this order ${\rm Tr}Z$ is undetermined. 
On the other hand, the next leading terms of order ${\cal O}(1/M_P^2)$ are given by
\bea
V_1 &\approx& -h\mu_2^2 {\rm Tr}(Z^\dag W + Z W^\dag) -3|W|^2 
+ (N_f-k)(2k \mu_1^2 + {\rm Tr}|Z|^2)|h \mu_2^2|^2
\nonumber \\
&=& - 2 h \mu_2^2 {\rm Tr}Z (-h\mu_2^2 {\rm Tr}Z + c) -3(-h\mu_2^2 {\rm Tr}Z + c)^2 
+ (N_f - k) (2\mu_1^2 + {\rm Tr}Z^2)(h\mu_2^2)^2 \nonumber \\
\eea
where $\langle Z \rangle = \langle Z^\dag \rangle$ is assumed for simplicity. 

Minimization condition 
\bea
0 &=& \frac{\partial V_1}{\partial {\rm Tr}Z} = - 2 h\mu_2^2 
\left[ (-2h\mu_2^2 {\rm Tr}Z + c) - 3(-h\mu_2^2 {\rm Tr}Z + c) +(N_f-k){\rm Tr}Z(h\mu_2^2) 
\right] \nonumber
\eea
determines its VEV as
\bea
\langle {\rm Tr}Z \rangle \approx \frac{2c}{(N_f-k+1)h\mu_2^2}. 
\label{trZ}
\eea
Note that the moduli is lifted by the constant superpotential $c$ 
and $U(1)_R$ symmetry is spontaneously broken, 
which is natural because the constant superpotential has an R-charge 2. 
This should be contrasted with the work \cite{GKK} 
in which the $U(1)_R$ explicit breaking term $\delta W = Z^2$ is added by hand 
and its overall coefficient must be fine-tuned to be very small to have a minimum. 
In our case, the fine tuning is reduced to the cancellation of the cosmological constant 
and it is a generic fine-tuning in the models based on supergravity. 

To cancel or obtain a tiny positive cosmological constant, 
we need the following fine tuning. 
\bea
0 &\approx& V_0 + V_1 \approx (N_f-k)(h \mu_2^2)^2 
-2\frac{2c^2}{N_f-k+1} \left( -\frac{2}{N_f-k+1} + 1 \right) 
\nonumber \\
&&-3 \left( -\frac{2}{N_f - k + 1} + 1 \right)^2 c^2 
+ (N_f-k)(h\mu_2^2)^2 \left[ 2k\mu_2^2 + \frac{4c^2}{(h\mu_2^2)^2(N_f-k+1)^2} \right]
\nonumber \\
\eea
which fixes the cosmological constant
\bea
c \approx \sqrt{\frac{N_f-k}{3(N_f-k-1)^2-4}}(N_f-k+1)h \mu_2^2.
\eea  
This leads to 
\bea
\langle {\rm Tr}Z \rangle &\approx& 2 \sqrt{\frac{N_f-k}{3(N_f-k-1)^2-4}}, \label{Z} \\
\langle W \rangle &\approx& (N_f-k-1)\sqrt{\frac{N_f-k}{3(N_f-k-1)^2-4}} h\mu_2^2.
\label{W} 
\eea
We should understand here the fact that the VEV of the moduli ${\rm Tr}Z$ 
and SUSY breaking scale $F_{{\rm Tr}Z}$ are determined by the constant superpotential $c$. 
Namely, if the vanishing constant superpotential limit $c \to 0$ is taken, 
the both of the $\langle {\rm Tr}Z \rangle$ and $\langle F_{{\rm Tr }Z} \rangle$ vanish. 
Therefore, the constant superpotential is essential in our model. 
\begin{figure}[h]
\begin{center}
\includegraphics[width=6cm]{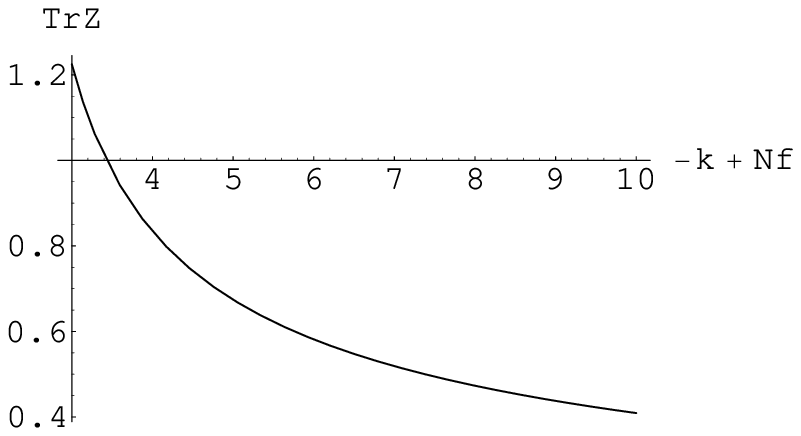}
\hspace*{8mm}
\includegraphics[width=6cm]{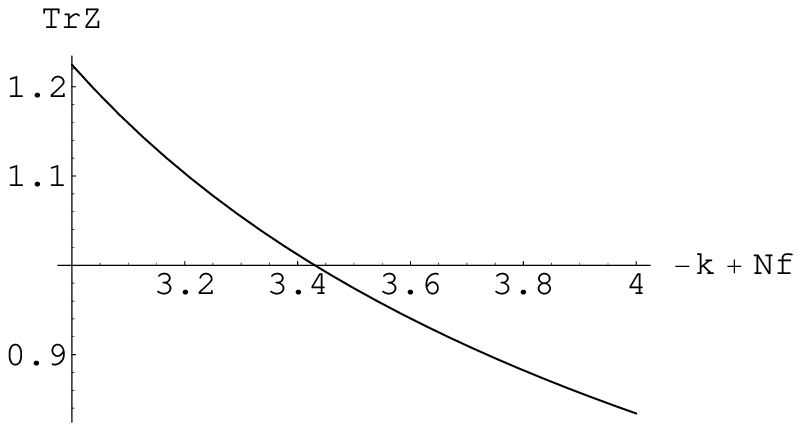}
\end{center}
\caption{The plot of $\langle {\rm Tr}Z \rangle$ as a function of $N_f-k$. 
The plot in the right-hand side is focused on the range 
around $\langle {\rm Tr}Z \rangle = 1$ of that in the left-hand side.}
\label{zN-k}
\end{figure}
From Fig.~\ref{zN-k}, for our analysis to be reliable in the context of effective theories,  
the moduli VEV must be smaller than the Planck scale $\langle {\rm Tr}Z \rangle < 1$, 
which is translated to a condition
\bea
N_f-k > 3.4. 
\eea
This constraint satisfies the assumption $k \le N = N_f-N_c$ if $N_c \ge 4$. 

We comment here on another vacuum with a lower energy 
than the vacuum discussed in this section. 
The fields with nonvanishing VEV in another vacuum are 
\bea
\chi_1, \tilde{\chi}_1 \simeq \mu_1, \quad 
\rho_1, \tilde{\rho}_1 \simeq \mu_2. 
\eea
The vacuum energy independent of Planck scale in this vacuum is found to be zero 
$(V_0)_{{\rm another}} = 0$. 
SUSY is broken (SUSY breaking scale is suppressed by Planck scale.). 
This vacuum is not preferable in the phenomenological viewpoint 
since the gaugino masses via gauge mediation are not generated 
because of vanishing VEV of the moduli $\langle Z \rangle=0$.\footnote{Note that 
R-symmetry is broken by the constant superpotential. 
In principle, the gravity mediated gaugino masses are of order 
$\langle W \rangle/M_P^2 \sim {\rm TeV}$ at tree level. 
However, as will be described later in the text, 
we assume throughout this paper that the hidden and the visible sectors are sequestered 
in the superspace density to avoid flavor dependent sfermion masses at tree level. 
As a result, the gravity mediated gaugino masses are not generated at tree level as well. 
Although the sizable gaugino masses can be obtained via anomaly mediation, 
we does not consider this possibility in this paper.} 
Therefore, the decay to this vacuum must be strongly suppressed 
so that the lifetime of our vacuum be longer than the age of the universe. 
Let us check that this is indeed the case. 
We note that the field distance $\Delta \Phi$ and the potential difference $\Delta V$ 
between our vacuum and another one are $\Delta \Phi \sim \langle {\rm Tr}Z \rangle$ and 
$\Delta V \sim -3c^2$. 
Therefore, the bounce action between them can be estimated as 
\bea
S_{{\rm bounce}} \sim \left| \frac{(\Delta \Phi)^4}{\Delta V} \right| 
\sim \frac{(\langle {\rm Tr}Z \rangle)^4}{3c^2} 
\sim \frac{(\langle {\rm Tr}Z \rangle)^4}{(h \mu_2^2)^2} 
\sim \frac{(\langle {\rm Tr}Z \rangle)^4}{(\langle F_{{\rm Tr}Z} \rangle)^2} 
\sim 10^{10-12} \gg {\cal O}(100)
\label{bounce}
\eea
where SUSY breaking scale $\langle F_{{\rm Tr}Z} \rangle \approx 10^{10-11}~{\rm GeV}^2$ 
will be determined later. 
(\ref{bounce}) means that the lifetime of our vacuum is 
longer than the age of universe. 
Thus, it ensures the stability of our vacuum. 

\section{Mass spectrum of messengers}

As will be discussed in the next section, 
if we embed the Standard Model gauge group 
into the gauged subgroup of global symmetry $SU(N_f-k)$ in the hidden sector, 
then the messengers are turned out to be 
$Z, Y, \tilde{Y}, \rho_{1,2}, \tilde{\rho}_{1,2}$. 
In this section, we solve the mass spectrum of messengers. 
The potential for messengers is  
\bea
V &=& e^K 
{\rm Tr}
\left[
|V^\dag W + h(\chi_1 \tilde{\chi}_1 - \mu_1^2)|^2 
+ |Z^\dag W +h (\rho_1 \tilde{\rho}_1 + \rho_2 \tilde{\rho}_2) - h \mu_2^2|^2 
\right. \nonumber \\
&& \left. 
+ |Y^\dag W + h (\chi_1 \tilde{\rho}_1)|^2 
+ |\tilde{Y}^\dag W + h (\tilde{\chi}_1 \rho_1)|^2 
\right. \nonumber \\
&& \left. 
+ |\chi_1^\dag W + h (V \tilde{\chi}_1 + Y \tilde{\rho}_1)|^2 
+ |\tilde{\chi}_1^\dag W + h (V \chi_1 + Y \rho_1)|^2 
\right. \nonumber \\
&& \left. 
+ |\rho_1^\dag W + h (\tilde{Y} \tilde{\chi}_1 + Z \tilde{\rho}_1)|^2 
+ |\tilde{\rho}_1^\dag W + h (Y \chi_1 + Z \rho_1)|^2 
\right. \nonumber \\
&& \left. 
+ |\rho_2^\dag W + h ( Z \tilde{\rho}_2)|^2 
+ |\tilde{\rho}_2^\dag W + h Z \rho_2|^2
+ |\chi_2^\dag W|^2 + |\tilde{\chi}_2^\dag W|^2 - 3|W|^2
\right]. 
\label{messpot}
\eea
The nonvanishing VEVs of the curvature of the potential are listed. 
\bea
\langle V_{ZZ^\dag} \rangle 
&=& e^{\langle K \rangle} 
\left[-h\mu_2^2 \langle Z^\dag W + Z W^\dag \rangle (-1+|\langle Z \rangle|^2) 
+3(h\mu_2^2)^2 |\langle Z \rangle|^2 -(h\mu_2^2)^2 
\right. \nonumber \\
&& \left. 
+(1+2|\langle Z \rangle|^2) |\langle W \rangle|^2 
+ (h\mu_2)^2 (|\langle \chi \rangle|^2 + |\langle \tilde{\chi} \rangle|^2) 
\langle Z + Z^\dag \rangle
\right] \nonumber \\
&\approx& -(h\mu_2^2)^2 - 4 h \mu_2^2 c \langle Z \rangle  + |\langle W \rangle|^2 
\approx (h\mu_2^2)^2, \\
\langle V_{YY^\dag} \rangle 
&=& \langle V_{\tilde{Y}\tilde{Y}^\dag} \rangle 
= e^{\langle K \rangle} 
\left[
(h \mu_1)^2 + |\langle W \rangle|^2
\right] \approx (h \mu_1)^2, \\
\langle V_{\rho_1 Y^\dag} \rangle 
&=& \langle V_{Y\rho_1^\dag} \rangle 
= e^{\langle K \rangle} h^2 \mu_1 \langle Z \rangle, \\
\langle V_{\tilde{\rho}_1 \tilde{Y}^\dag} \rangle 
&=& \langle V_{\tilde{Y} \tilde{\rho}_1^\dag} \rangle 
= e^{\langle K \rangle} h^2 \mu_1 \langle Z \rangle^\dag, \\ 
\langle V_{\rho_1 \rho_1^\dag} \rangle 
&=& \langle V_{\tilde{\rho}_1 \tilde{\rho}_1^\dag} \rangle
= e^{\langle K \rangle} [(h \langle Z \rangle)^2 + |\langle W \rangle|^2 + (h \mu_1)^2] 
\approx e^{\langle K \rangle} (h \langle Z \rangle)^2, \\
\langle V_{\rho_2 \rho_2^\dag} \rangle &=& 
\langle V_{\tilde{\rho}_2 \tilde{\rho}_2^\dag} \rangle  
= e^{\langle K \rangle} [(h \langle Z \rangle)^2 + |\langle W \rangle|^2 ] 
\approx e^{\langle K \rangle} (h \langle Z \rangle)^2. 
\eea
In this estimation, 
the terms suppressed by the Planck scale are simply neglected 
or $h \langle Z \rangle \gg h \mu_1$ is considered. 
It is easy to find the masses of $Z, \rho_2, \tilde{\rho}_2$ 
since they have no mixing.  
If we take into account SUSY breaking mass for $\rho_{1,2}, \tilde{\rho}_{1,2}$, 
the mixing takes place from the terms
\bea
V \supset h F_{{\rm Tr}Z} (\rho_1 \tilde{\rho}_1 + \rho_2 \tilde{\rho}_2) + {\rm h.c.}, 
\eea
then the mass terms of $\rho_2, \tilde{\rho}_2$ sector are read off
\bea
&&(\rho_2^\dag, \tilde{\rho}_2)
\left(
\begin{array}{cc}
|M|^2 & (h F_{{\rm Tr}Z})^\dag \\
(h F_{{\rm Tr}Z}) & |M|^2 \\
\end{array}
\right)
\left(
\begin{array}{c}
\rho_2 \\
\tilde{\rho}_2^\dag \\
\end{array}
\right)
\eea
where $|M|^2 \equiv e^{\langle K \rangle}|\langle W \rangle|^2 + |h \langle {\rm Tr}Z \rangle|^2$. 
The eigenvalues are
\bea
m^2_{\rho^\pm_2} = e^{\langle K \rangle} [|M|^2 \pm h \langle F_{{\rm Tr}Z} \rangle ] 
\approx (h \langle {\rm Tr}Z \rangle)^2 
\approx {\cal O}((0.1M_P)^2)
\eea
and the corresponding eigenstates are 
$\rho_2^\pm \equiv (\rho_2 \pm \tilde{\rho}_2^\dag)/\sqrt{2}$. 
The approximation $(h\langle Z \rangle)^2 \gg \langle F_{{\rm Tr}Z} \rangle$ 
is taken, which will be justified later. 

The mass matrix for $Y, \tilde{Y}, \rho_1, \tilde{\rho}_1$ before SUSY breaking
is found to be
\bea
(Y^\dag, \rho_1^\dag, \tilde{Y}^\dag, \tilde{\rho}_1^\dag)
\left(
\begin{array}{cccc}
m_{|Y|^2}^2 & m^2_{Y^\dag \rho_1} & 0 & 0 \\
m_{\rho_1^\dag Y}^2 & m^2_{|\rho_1|^2} & 0 & 0 \\
0 & 0 & m_{|\tilde{Y}|^2}^2 & m^2_{\tilde{Y}^\dag \tilde{\rho}_1} \\
0 & 0 & m_{\tilde{Y} \tilde{\rho}_1^\dag}^2 & m^2_{|\tilde{\rho}_1|^2} \\
\end{array}
\right)
\left(
\begin{array}{c}
Y \\
\rho_1 \\
\tilde{Y} \\
\tilde{\rho}_1 \\
\end{array}
\right)
\eea
where
\bea
m^2_{|Y|^2} = m^2_{|\tilde{Y}|^2} \approx (h \mu_1)^2, \quad 
m^2_{Y^\dag \rho_1} = m^2_{\tilde{Y}^\dag \tilde{\rho}_1} 
\approx h^2 \mu_1 \langle {\rm Tr}Z \rangle, \quad
m^2_{|\rho_1|^2} = m^2_{|\tilde{\rho}_1|^2} \approx (h \langle {\rm Tr}Z \rangle)^2. 
\nonumber \\
\eea

Diagonalizing the mass matrix in $Y$-$\rho_1$ sector, 
we obtain eigenvalues 
$(h \langle {\rm Tr}Z \rangle)^2~{\rm or}~(h \mu_1)^2$
and the corresponding eigenmodes are 
\bea
\hat{\tilde{\rho}}_1 \equiv \tilde{\rho}_1 + \frac{\mu_1}{\langle {\rm Tr}Z \rangle} \tilde{Y}, \quad  
\hat{\tilde{Y}} \equiv \tilde{Y} - \frac{\mu_1}{\langle {\rm Tr}Z \rangle} \tilde{\rho}_1. 
\eea
In the above calculation, 
we assumed no accidental cancellation among ${\cal O}(1)$ coefficients. 
A similar calculation for $\tilde{Y}$-$\tilde{\rho}_1$ sector is also applied. 
Taking into account SUSY breaking for $\rho_1, \tilde{\rho}_1$ sector, 
the leading order mass matrix for $\hat{\rho}_1, \hat{\tilde{\rho}}_1$ sector 
is identical to the one for $\rho_2, \tilde{\rho}_2$ sector. 
Therefore, the mass eigenvalues and its corresponding mass eigenstates are
\bea
&&m_{\hat{\rho}_1^\pm} = e^{\langle K \rangle} [(h \langle {\rm Tr}Z \rangle)^2 
\pm h \langle F_{{\rm Tr}Z} \rangle]
\approx (h \langle {\rm Tr}Z \rangle)^2, \\
&&\hat{\rho}_1^\pm \equiv (\hat{\rho}_1 \pm \hat{\tilde{\rho}}_1^\dag)/\sqrt{2}. 
\eea
On the other hand, 
the masses of messenger fermions are read from the superpotential \cite{GKK}
\bea
W = -h\mu_2^2 {\rm Tr}Z + h \sum_{i=1}^{N-k} {\rm Tr}Z \rho_2^i (\tilde{\rho}_2)_i 
+ h \sum_{i=1}^k \left[ {\rm Tr}Z \rho_1^i (\tilde{\rho}_1)_i 
+ \mu_1 \rho_1^i \tilde{Y}_i + \mu_1 Y^i (\tilde{\rho}_1)_i \right]
\label{messsp}
\eea
as
\bea
h(\psi_{\tilde{\rho}_1}, \psi_{\tilde{Y}}, \psi_{\tilde{\rho}_2})
\underbrace{\left(
\begin{array}{ccc}
{\rm Tr}Z & \mu_1 & 0 \\
\mu_1 & 0 & 0 \\
0 & 0 & {\rm Tr}Z \\
\end{array}
\right)}_{\equiv {\cal M}}
\left(
\begin{array}{c}
\psi_{\rho_1} \\
\psi_{Y} \\
\psi_{\rho_2} \\
\end{array}
\right). 
\label{M}
\eea
$\psi_{\rho_2}$ is decouple and 
the mass eigenvalues of $\rho_1$-$Y$ sector is obtained 
as $h \langle {\rm Tr}Z \rangle$ and $h\mu_1$. 
We see that some (the other) linear combination of messenger multiplets 
in $\rho_1$-$Y$ sector 
is heavy (light) with mass $h \langle {\rm Tr}Z \rangle (h \mu_1)$. 

Note that the fermionic component of $Z$ in the adjoint representation is massless at tree level. 
However, it will be massive at one-loop level through the ordinary gauge-mediated diagram.
\bea
m_{\psi_Z} \simeq \frac{\alpha_h}{4\pi} \frac{\langle F_{{\rm Tr}Z} \rangle}{\langle {\rm Tr}Z \rangle} 
\sim 0.01 \times \frac{(10^{11}~{\rm GeV})^2}{0.1M_P~{\rm GeV}} \sim {\cal O}({\rm TeV}). 
\eea
Trace part of fermionic component of $Z$ corresponds to NG fermion, 
which is absorbed into the longitudinal component of the gravitino after SUSY breaking.

\section{Direct Gauge Mediation}

Gauging a subgroup $SU(N_f-k)$ of the unbroken flavor symmetry $SU(k) \times SU(N_f-k)$ 
of the model, SUSY breaking is transmitted to the MSSM 
through the messenger $Y, \tilde{Y}, \rho_{1,2}, \tilde{\rho}_{1,2}$ loops. 
In the uplifted ISS model, the superpotential for messengers takes of the form (\ref{messsp}). 
The second term in the $\rho_2, \tilde{\rho}_2$ sector 
corresponds to the messenger interactions in the minimal gauge mediation, 
which gives contributions to {\em both} gaugino and sfermion masses. 
However, the last three terms in the $Y$-$\rho_1$ sector provide 
contributions to the sfermion masses only as in the ISS model. 

Gaugino and sfermion masses are calculated in \cite{CFS}. 
\bea
M_r &=& \frac{\alpha_r}{4\pi}\Lambda_G, \\
m^2 &=& 2 \sum_{r=1}^3 C_r \left(\frac{\alpha_r}{4\pi} \right)^2 \Lambda^2_S
\eea
where $r=1,2,3$ standing for $U(1), SU(2), SU(3)$, respectively. 
$C_r$ is the quadratic Casimir in the gauge group corresponding to $r$. 
\bea
&&\Lambda_G \equiv \langle F_{{\rm Tr}Z} \rangle \partial_{{\rm Tr}Z} \log \det {\cal M}, \\
&&\Lambda_S^2 \equiv \frac{1}{2} |\langle F_{{\rm Tr}Z} \rangle|^2 
\frac{\partial^2}{\partial {\rm Tr}Z^\dag \partial {\rm Tr}Z} 
\sum_{i=1}^N \left(\log |{\cal M}|^2 \right)^2. 
\eea
$Z$ dependent mass matrix ${\cal M}$ is given in (\ref{M}). 
Using this ${\cal M}$, we obtain the gaugino and sfermion masses 
\bea
M_r &=& \frac{\alpha_r}{4\pi}(N-k) \frac{\langle F_{{\rm Z}} \rangle}{\langle {\rm Tr}Z \rangle} 
\approx \frac{\alpha_r}{4\pi}(N-k)\frac{h\mu_2^2}{0.1M_P}, \\
m^2 &=& 2 \sum_{r=1}^3 C_r \left(\frac{\alpha_r}{4\pi} \right)^2 |\langle F_{{\rm Tr}Z} \rangle|^2 
\times \nonumber \\
&&\left[
\frac{N-k}{|\langle {\rm Tr}Z \rangle|^2} + \frac{2k}{|\langle {\rm Tr}Z \rangle|^2 + 4 \mu_1^2} 
+ \frac{ 2k \log \left( \frac{ 
|\langle {\rm Tr}Z \rangle|^2 + 2\mu_1^2 + |\langle {\rm Tr}Z \rangle| 
\sqrt{|\langle {\rm Tr}Z \rangle|^2 + 4 \mu_1^2}}
{ |\langle {\rm Tr}Z \rangle|^2 + 2\mu_1^2 - |\langle {\rm Tr}Z \rangle| 
\sqrt{|\langle {\rm Tr}Z \rangle|^2 + 4 \mu_1^2}} \right)}
{(|\langle {\rm Tr}Z \rangle|^2 + 4 \mu_1^2) \sqrt{|\langle {\rm Tr}Z \rangle|^4 
+ 4 \mu_1^2 |\langle {\rm Tr}Z \rangle|^2}}
\right] \nonumber \\
&\approx& 2 \sum_{r=1}^3 C_r \left(\frac{\alpha_r}{4\pi} \right)^2
(N+k) \left(\frac{h\mu_2^2}{\langle {\rm Tr}Z \rangle} \right)^2 
\approx \left(\frac{\alpha_r}{4\pi} \right)^2
\left( \frac{h\mu_2^2}{0.1M_P} \right)^2 
\eea
where $\langle {\rm Tr}Z \rangle \gg \mu_1$ are taking into account in the sfermion masses. 
Here we note that the gaugino and the sfermion masses are comparable 
(up to ${\cal O}(1)$ constant) similar to the ordinary gauge mediation case. 

Requiring $M_r \simeq m$ to be 1 TeV, we obtain
\bea
(N-k)h\mu_2^2 \approx 10^4 M_P \to 
\langle F_{{\rm {\rm Tr}Z}} \rangle = h \mu_2^2 \approx (10^{10 \sim 11}{\rm GeV})^2.  
\eea
This leads to the gravitino mass as
\bea
m_{3/2} = \frac{\langle F_{{\rm Tr}Z} \rangle}{\sqrt{3}M_P} 
\approx \frac{10^4M_P}{\sqrt{3}M_P} \simeq 1~{\rm TeV}.  
\eea
This result means that the gravity mediation effects to the sfermion masses 
are comparable to the gauge mediation one. 
This is an undesirable feature since the gravity mediation to the sfermion masses 
are generated at tree level from the flavor dependent contact term 
$\int d^4\theta Z^\dag Z Q^\dag_i Q_j$($i,j$:~flavor indices, 
$Q$:~the MSSM multiplets) 
and cause the large flavor violation at unacceptable level in general. 
Therefore, we need to ``sequester" the hidden sector from the visible sector at tree level. 
This can be realized if the hidden sector and the visible sector are separated 
in the superspace density $\varphi$ not in the K\"ahler potential \cite{IKYY}. 
Namely, 
\bea
\varphi = \varphi_h(Z^\dag, Z) + \varphi_v(Q^\dag, Q). 
\label{sequest}
\eea
The superspace density is related to the K\"ahler potential as $K=-3\ln \varphi$.  
As a result, the gravity mediated sfermion masses vanish at tree level and 
the next leading masses are generated by anomaly mediation 
if the compensator multiplet has an F-term VEV. 
Throughout this paper, we assume the form of the superspace density (\ref{sequest}) 
and the compensator does not develop F-term VEV for simplicity 
because our focus is on the direct gauge mediation.

\section{Landau pole analysis}

In this model, there are extra multiplets with the SM charges ({\em i.e.} messengers) 
other than the MSSM multiplets, 
namely adjoint representation of $Z$, 
and the (anti) fundamental ones $Y, \tilde{Y}, \rho_{1,2}, \tilde{\rho}_{1,2}$. 
As was calculated before, the masses of $\rho_2, \tilde{\rho}_2$ and 
$\rho_1 + \frac{h \mu_1}{h \langle Z \rangle}Y, 
\tilde{\rho}_1 + \frac{h \mu_1}{h \langle Z \rangle} \tilde{Y}$ 
are ${\cal O}(0.1M_P) > M_{{\rm GUT}}$, 
so these multiplets do not affect the gauge coupling unification. 
On the other hand, the multiplets $Y - \frac{h \mu_1}{h \langle Z \rangle}\rho_1, 
\tilde{Y} - \frac{h \mu_1}{h \langle Z \rangle} \tilde{\rho}_1$ have masses of order $h\mu_1$. 
$Z$ have a mass of order ${\cal O}$(TeV), which have to be taken into account 
in the gauge coupling running. 
As will be shown below, 
the Landau pole problem for QCD coupling can be easily avoided 
comparing to the model of \cite{GKK} since more than half of messengers become superheavy 
of order $0.1M_P$.

The mass spectrum is summarized below. 
\bea
&&m_{\hat{\rho}_1}, m_{\hat{\tilde{\rho}}_1}, m_{\rho_2}, m_{\tilde{\rho}_2} \approx 0.1M_P, 
\nonumber \\ 
&&m_{\hat{Y}}, m_{\hat{\tilde{Y}}} \approx h \mu_1, \quad
m_\lambda \simeq m_{\tilde{f}} \approx m_Z \approx 1~{\rm TeV}. 
\eea
Now, we study a Landau pole constraint for QCD gauge coupling.  
We consider the two cases where the messengers are embedded into the MSSM multiplets 
and the SU(5) GUT multiplets. 

One-loop gauge coupling RGE is given by 
\bea
g_i^{-2}(\mu) = g_i^{-2}(\mu') + \frac{b_i}{8\pi^2}
\ln \left(\frac{\mu}{\mu'} \right)
\eea
where $\mu, \mu'$ are the renormalization scales. 
$b_i$ is one-loop beta function coefficient of the gauge group $i=SU(3), SU(2), U(1)$.

First, we study the MSSM embedding case. 
The one-loop beta function coefficients for QCD coupling 
at various scales are listed below. 
\bea
m_W < \mu < {\rm TeV} &:& b_3 = b_3^{{\rm SM}} = 7 \nonumber \\
{\rm TeV} < \mu < h\mu_1 &:& b_3 = b_3^{{\rm MSSM}} - \underbrace{b_3(Z)}_{3} = 0 \nonumber \\
h \mu_1 < \mu < h \langle Z \rangle \approx {\cal O}(0.1M_P) &:& b_3 
= b_3^{{\rm MSSM}} - b_3(Z) - \underbrace{b_3(\hat{Y}, \hat{\tilde{Y}})}_{k \times (1/2+1/2)} =- k
\eea
where $b_3^{{\rm SM}}$, $b_3^{{\rm MSSM}}$ are the QCD one-loop beta function 
coefficients for the standard model and the minimal SUSY standard model. 

The requirement that the QCD coupling at the GUT scale is perturbative 
\bea
\alpha_3(M_{{\rm GUT}}) = \frac{1}{\alpha_3(m_W)^{-1}
-\frac{1}{2\pi}[7\ln(m_W/{\rm TeV}) - k \ln(h\mu_1/M_{{\rm GUT}})]} < 1 
\eea
constrains the number of messengers $k$ and the mass scale $h\mu_1$ as
\bea
x > 16-\frac{27.5}{k}. 
\label{MSSM}
\eea
where we made use of $\alpha_3(m_W) \simeq 0.118$ and 
assumed $h\mu_1 \equiv 10^x$ GeV to be smaller than the GUT scale.

Similarly, noting that extra multiplets $({\bf 3,2}), ({\bf 3^*,2})$ 
with the SM charges from $Z$ multiplet 
contribute to the QCD gauge coupling running in the GUT embedding case, 
we obtain a similar constraint
\bea
x > 16-\frac{14.5}{k}. 
\label{GUT}
\eea 
\begin{figure}[h]
\begin{center}
\includegraphics[width=6cm]{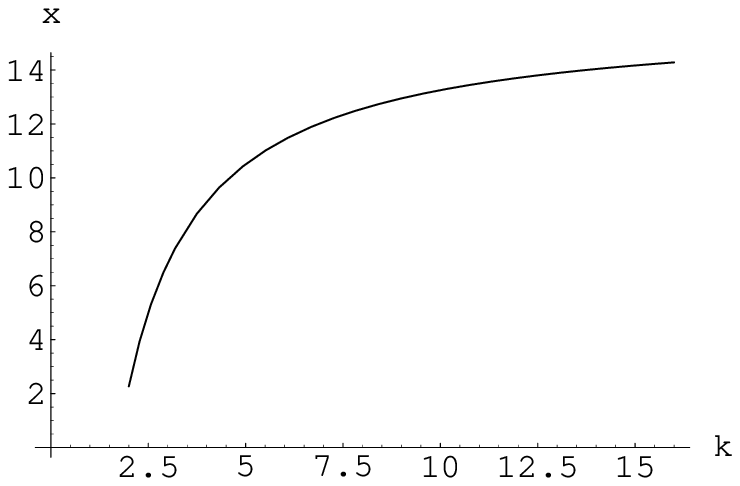}
\hspace*{8mm}
\includegraphics[width=6cm]{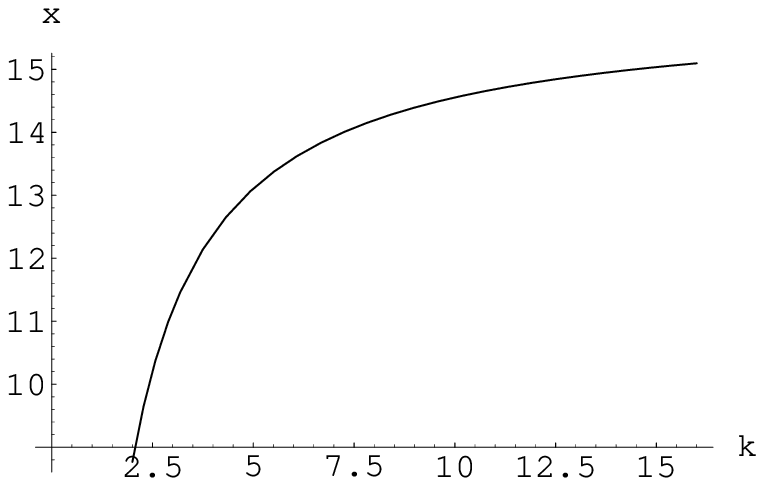}
\end{center}
\caption{The plot of constraints for the mass scale $h\mu_1$ 
as a function of the number of messengers $k$, 
the MSSM case (\ref{MSSM})(left) and the GUT case (\ref{GUT})(right).}
\label{Landau}
\end{figure}
As an illustration, we show plots of (\ref{MSSM}) and (\ref{GUT}) in Fig. \ref{Landau}.  
Here we assume the mass scale $h \mu_1$ to be larger than TeV 
since the messengers $\hat{Y}, \hat{\tilde{Y}}$ have not been observed below the TeV scale. 
The upper region of the plot implies that the perturbative unification is possible. 
More concretely, some numerical examples allowed by corresponding constraints 
are listed in the following Table. 
\bea
\begin{array}{|c|c|c|}
\hline
k & x({\rm MSSM}) & x({\rm GUT})\\
\hline
1 & {\rm No~constraint} & {\rm No~constraint} \\
\hline
2 & {\rm No~constraint} & > 8.77 \\
\hline
3 & > 6.84 & > 11.1 \\
\hline
4 & > 9.13 & > 12.4 \\
\hline
\end{array}
\eea
If $h\mu_1 > M_{{\rm GUT}}$, only the $Z$ multiplet in the adjoint representation 
contributes to the QCD gauge coupling running. 
In this case, we can easily check that the perturbative unification is possible 
without any problem. 

\section{Summary}

In this paper, we have discussed a model of direct gauge mediation based on 
an uplifted metastable SUSY breaking model in supergravity. 
The constant superpotential plays an essential role to fix the moduli 
as well as breaking SUSY and R-symmetry and the cancellation of the cosmological constant. 
SUSY breaking is directly gauge mediated to superparticles 
by gauging a subgroup of unbroken flavor symmetry. 
In particular, the gaugino masses are generated at leading order of SUSY breaking scale, 
namely comparable to the sfermion masses as in the ordinary gauge mediation. 
Improvements by simply coupling an uplifted metastable SUSY breaking model \cite{GKK} 
to supergravity are two fold. 
One is that the ad hoc fine-tuning to fix the moduli in the model of \cite{GKK} 
reduces to the fine-tuning to cancel the cosmological constant 
necessary and generic for any model based on supergravity
The other is that Landau pole problem for QCD coupling can be easily solved 
since more than half of messengers become superheavy, 
which are heavier than the GUT scale. 
The constraint for the number of messengers and their mass scale is obtained 
by the effects of the remaining light messengers to the running of QCD coupling constant. 

In our analysis, the anomaly mediation contribution is simply neglected 
by assuming a vanishing F-term of the compensator 
since our main interests are focused on the direct gauge mediation. 
Relaxing this assumption, the pattern of superparticle spectrum would become quite rich 
by the interplay of gauge and anomaly mediations. 
Studying such a spectrum in detail will be left for a future work. 

 \subsection*{Acknowledgments}
The work of the author was supported 
 in part by the Grant-in-Aid for Scientific Research 
 of the Ministry of Education, Science and Culture, No.18204024.


\end{document}